# Comparing Radio Propagation Channels Between 28 and 140 GHz Bands in a Shopping Mall

Sinh L. H. Nguyen[†], Jan Järveläinen[†‡], Aki Karttunen[†], Katsuyuki Haneda[†], and Jyri Putkonen[*]

[†]Aalto University, School of Electrical and Engineering, Finland
{sinh.nguyen, katsuyuki.haneda, aki.karttunen}@aalto.fi

[‡]Premix Oy, Finland
jan.jarvelainen@premixgroup.com

[*]Nokia Bell Labs, Finland
jyri.putkonen@nokia-bell-labs.com

*Abstract*—In this paper, we compare the radio propagation channels characteristics between 28 and 140 GHz bands based on the wideband (several GHz) and directional channel sounding in a shopping mall environment. The measurements and data processing are conducted in such a way to meet requirements for a fair comparison of large- and small- scale channel parameters between the two bands. Our results reveal that there is high spatial-temporal correlation between 28 and 140 GHz channels, similar numbers of strong multipath components, and only small variations in the large-scale parameters between the two bands. Furthermore, when including the weak paths there are higher total numbers of clusters and paths in 28 GHz as compared to those in 140 GHz bands. With these similarities, it would be very interesting to investigate the potentials of using 140 GHz band in the future mobile radio communications.

## I. INTRODUCTION

Millimeter-wave bands, referred as frequency bands from 30 GHz−300 GHz, have been exploited for 5G radio communications in recent years [1], [2]. However, the majority of the propagation channel studies and experiments has been focused only to bands up to 100 GHz [3]–[7]. While above-100 GHz bands experience higher path loss, at the same time wider unused bandwidth chunks are available in those bands, making them also possible candidates for future wireless systems. For example, in Finland a bandwidth of 7.5 GHz (141 − 148.5 GHz) in D-band is currently unused and could be exploited for future fixed and mobile radio communications [8].

Shopping mall is among the scenarios where mobile broadband experiences are expected to be supported in 5G, i.e., "great service in a crowd" [9]. Propagation channels in the shopping mall environment were studied in [10] at 28 GHz band, and in [11], [12] for 15, 28, 60, and 73 GHz bands. Literature on the channel measurements at above 100 GHz frequencies is in general very limited, and has not been found for the shopping mall environment in particular. Only one previous work on very short range indoor pathloss measurement has been found in the literature, where line-of-sight (LOS) pathloss was measured as the antenna separation varied from 20 to 180 cm [13].

In this paper, for the first time in the literature we report *microcell directional channel measurements* at 140 GHz for a large indoor shopping mall environment. Furthermore, we compare the 140 GHz channel characteristics with its counterparts at 28 GHz using the data measured at the same links in the same manner. The wideband (4 GHz) and directional measurements and data processing were conducted in such a way allowing us to make a fair comparison of large- and small- scale channel parameters between the two bands. The similarity and difference in pathloss, and large- and small-scale parameters between the two bands are then analyzed to investigate the possibility of using the 140 GHz frequencies for future mobile radio communications.

## II. MEASUREMENT SETUP AND ENVIRONMENT

The wideband directional channel measurements were performed using the same approach reported in [11], [14], [15]. Specifically, two similar setup radio channel sounders were used to perform channel measurements in the 28 and 140 GHz frequency ranges, where the RF signals were generated using the Ka-band (26.5 − 40 GHz) and D-band (110 − 170 GHz) up- and down-converters, respectively. The schematic diagram of the measurement system is shown in Fig. 1. The details of the channel sounder can be found in [11].

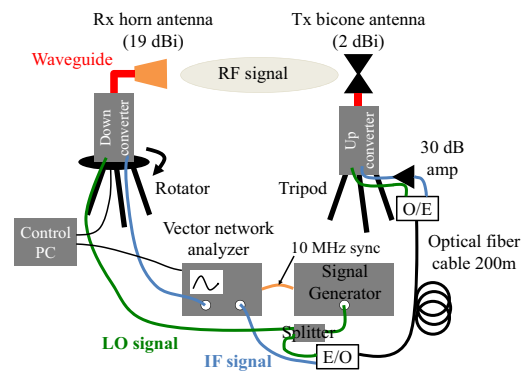

Fig. 1. Channel sounding system: Tx (Rx) includes a frequency multiplier (factor of 2 in 28-GHz and 12 in 140-GHz bands) and a mixer for up-converting (down-converting) the IF (RF) signal.

The venue for the shopping mall measurements was the Sello shopping mall in Leppävaara, Espoo, Finland, presented

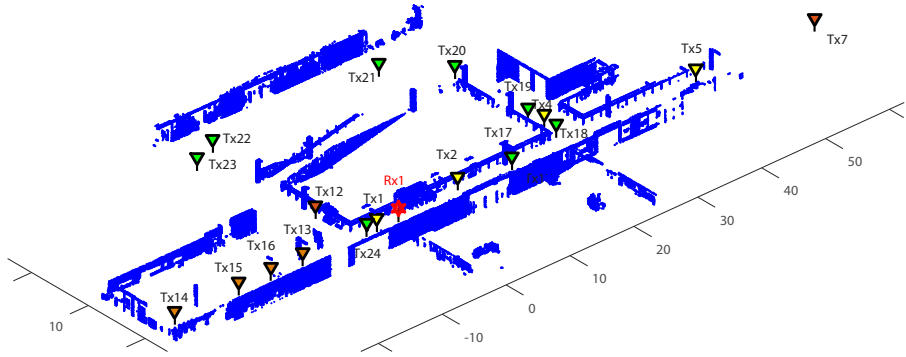

Fig. 2. 3-D map of the Tx and Rx positions in the 3rd floor of the Sello shopping mall in the 140-GHz measurement, overlaid with the 3-D point cloud model of the environment. The green triangles present the Tx locations that were also measured at 28 GHz in the same day in November 2016; the yellow triangles present the Tx locations that were also measured at 28 GHz but in March 2015; the orange triangles present the Tx locations that were measured in 140 GHz band only.

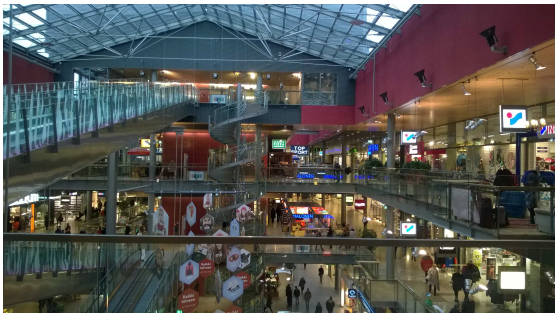

Fig. 3. Photo from the Sello shopping mall.

TABLE I
SUMMARY OF THE MEASUREMENTS.

| Parameter | 28-GHz band | 140-GHz band |
|---|---|---|
| Center frequency | 28.5 GHz | 143.1 GHz |
| Bandwidth | 4 GHz | 4 GHz |
| Transmit power | 2 dBm | $-7$ dBm |
| PDP dynamic range | 123 dB | 130 dB |
| Tx / Rx height | 1.9 /1.9 m | |
| Link distance range | 3 − 65 m | |
| Rx antenna | horn, 19 dBi, $10°$ az./$40°$ el. HPBW | |
| Tx antenna | bicone, 0 dBi, $60°$ el. HPBW | |

measurements in [11], [14].

To compare 140 GHz channels with its 28 GHz counterpart, 8 links (5 LOS and 3 obstructed LOS) that were measured in the same day are used for the comparison. As can be seen from Table II and Fig. 4, measurement specifications including bandwidth and antenna patterns were ensured to meet the requirements for comparability of channel's parameters across different frequencies, as defined in [6].

TABLE II
COMPARABILITY BETWEEN TWO FREQUENCY BANDS.

| Requirement | Comment |
|---|---|
| Same environment | ✓ |
| Same measurement time | Approximately (in the same day) |
| Same antenna locations | ✓ |
| Comparable antenna patterns | ✓ |
| Equal delay resolution | 0.25 ns |
| Equal spatial resolution | $10°$ in azimuth, $40°$ in elevation |
| Same post processing method | ✓ |
| Same power range | ✓ |

in Fig. 3. The shopping mall is a modern, four-story building with a large open space in the middle and approximate dimensions of $120 \times 70$ m$^2$. The floorplan of the channel measurements are shown in Fig. 2. In total, 18 Tx antenna locations were measured at 140 GHz, with antenna heights of 1.9 m and the Tx-Rx distance ranging from 3 to 65 m, approximately. The Tx were moved a long the corridor and around open space in the middle of the shopping mall. In three Tx locations, the LOS path was obstructed by the pillar or the escalator. The antenna locations were chosen such that human blockage could be avoided in order to maintain the repeatability of the measurements at both frequency bands. At each Tx location, the Rx horn antenna was rotated in the azimuth plane with $5°$ step, as similarly done in the directional

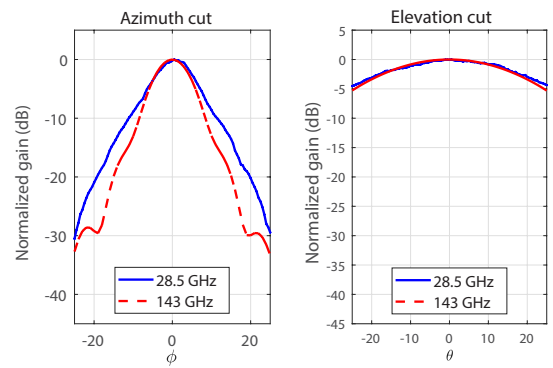

Fig. 4. Comparison of Rx radiation patterns in (left) azimuth plane and (right) elevation plane between 28-GHz and 140-GHz bands.

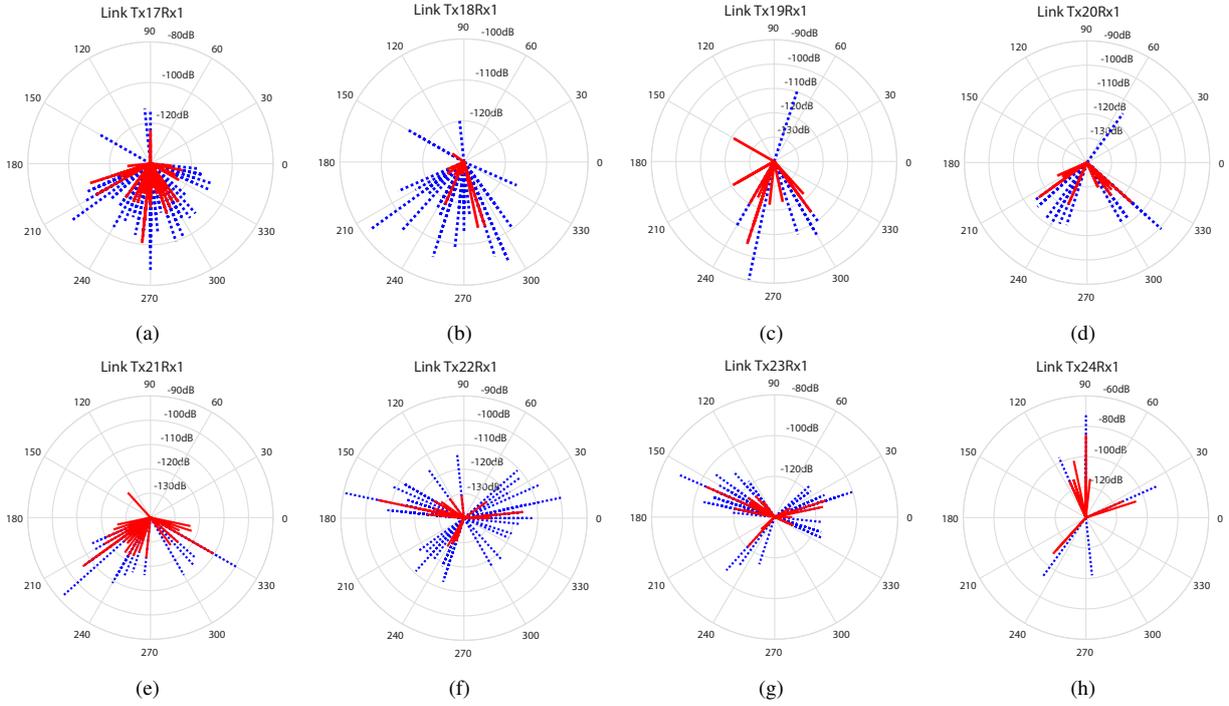

Fig. 5. Comparison of power-angular profiles between 28 GHz (dashed blue) and 140 GHz (solid red) for Tx positions from ((a) 17 to (h) 24. Tx positions 17, 19, 21, 23, 24 are in LOS and the rest are in obstructed LOS conditions.

## III. COMPARISON OF LARGE-SCALE PARAMETERS

From the measured power angular delay profiles (PADPs), the multipath components (MPCs) in each measurement were extracted using the peak search algorithm in [14]. The peak detection algorithm is based on the assumption that the channel is *deterministic* for at least the strong MPCs, i.e., the arriving signals from discrete specular reflectors are completely resolvable either in delay or angular domain [16]. The assumption is justified by the very wide channel measurement bandwidth of 4 GHz and narrow azimuth beamwidth of $10°$ of the receive horn antennas at both measured frequencies. The azimuth angle of arrival (AoA) and amplitude of each MPC are calculated using the receive antenna pattern and subtracting the corresponding antenna gain from the peak's amplitude [14].

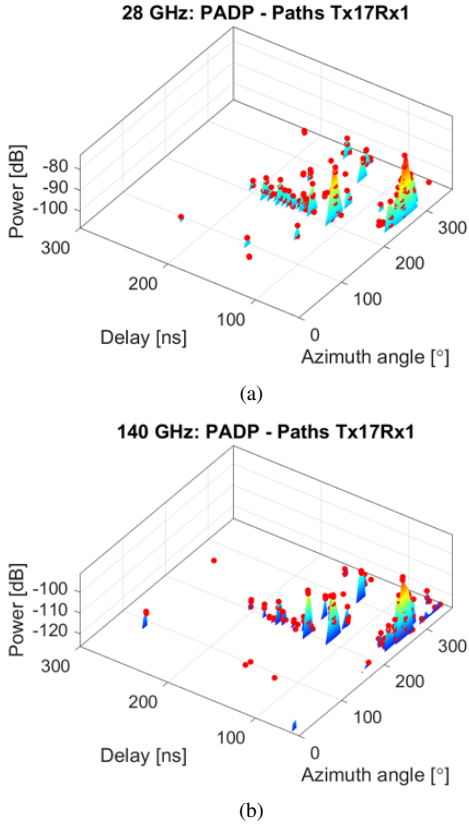

Fig. 6. PADPs of link Tx17-Rx1 and detected peaks (red dots) within 30-dB range in (a) 28 GHz and (b) 140 GHz measurements.

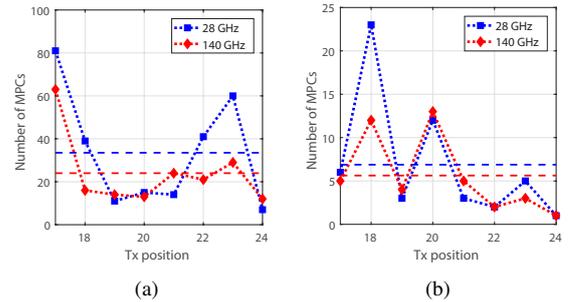

Fig. 7. Number of detected specular paths within a) 30-dB threshold and b) 15-dB threshold in Sello shopping mall. The corresponding mean values plotted with horizontal dotted lines

One example of the PADPs and peak search is presented in Fig. 6, which shows the measured channel for the Tx position 17 (LOS). It can be noticed that many deterministic paths occur in both frequency bands. As depicted in Figs. 5 and 7, there are clearly more paths at 28 GHz than at 140 GHz when considering a 30-dB threshold seen from the strongest path amplitude. However, when 15-dB threshold is used, the number of (strong) paths is very similar between the two bands for all links, expect link Tx18-Rx1, which is OLOS. This similarity in the multipath richness between the two bands, especially in this indoor environment, can be explained by the fact that the environment consists of many surfaces considered smooth even at 140 GHz.

From the detected MPCs, omni-direction pathloss, delay and angular spreads have been calculated for 28 and 140 GHz. Due to the low dynamic range at 140 GHz, a 30 dB threshold has been used. The results are presented in Figs. 8 and 9, and Table IV. It can be seen that the average delay spread is almost identical for the both frequency bands, and the azimuth spread is 10% higher at 28 GHz. The small difference between the bands can be explained by the fact that the most significant paths are found in both bands. When comparing link by link, noticeably higher DS and AS can be observed in 28 GHz in the obstructed LOS link Tx18-Rx1.

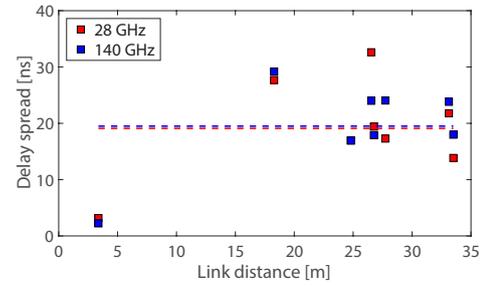
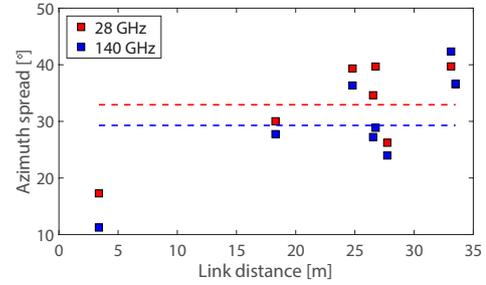

Fig. 9. (a) Delay spread and (b) azimuth spread in the shopping mall. Mean values plotted with dotted lines.

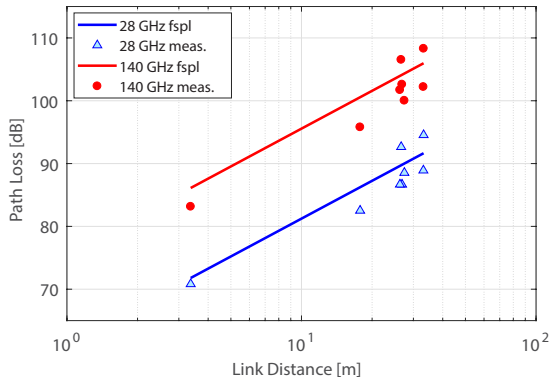

Fig. 8. Measured omni-direction pathloss at 28 GHz (dots) and 140 GHz (triangles) bands. The solid lines show the free-space pathloss (fspl) at corresponding frequencies.

Fitting the measured omni-directional pathloss data of each of the two bands to the pathloss model equation

$$PL(d) = 10A\log_{10}(d/1\text{m}) + B + N(0, \sigma^2), \quad (1)$$

we obtain the model parameters shown in Table III. It can be seen from Fig. 8 and Table III that except some additional fspl at 140 GHz, the slope and variations of the pathloss data of two bands are similar.

TABLE III
PATHLOSS MODEL PARAMETERS IN THE SHOPPING MALL IN 28-GHZ AND 140-GHZ BANDS.

| Parameter | 28 GHz | 140 GHz |
|---|---|---|
| $A$ | 2.10 | 2.22 |
| $B$ | 59.16 | 70.77 |
| $\sigma$ | 2.85 | 2.94 |

TABLE IV
MEAN OF DELAY AND AZIMUTH SPREADS OVER ALL COMMON TX-RX LINKS IN THE SHOPPING MALL AT 28 AND 140 GHZ.

| Frequency | Delay spread [ns] | Azimuth spread [°] |
|---|---|---|
| **28 GHz** | 19 | 33 |
| **140 GHz** | 19 | 29 |

## IV. COMPARING NUMBER OF CLUSTERS AND INTER-CLUSTER CHARACTERISTICS

To obtain cluster parameters for each link of the two bands, a hierarchical clustering algorithm [14] was used with the detected MPCs. For the purpose of comparing cluster model parameters between the two bands, the same multipath component distance (MCD) threshold of 30 dB was used for both 28 and 140 GHz band. Denoting the number of clusters and the number of MPCs per cluster for a given link as $N$ and $M$, respectively, Table V presents the mean and standard deviation values of $N$ and $M$ in the 28 and 140 GHz measurements. It can be observed from the results that both the number of clusters and the number of paths per cluster are higher in the 28 GHz bands as expected from the higher total number of detected paths when higher threshold value is used. Comparing to the parameters adopted in 3GPP New Radio (NR) model TR 38.901 for above 6 GHz Indoor Office environment [4], that is, 15 clusters for LOS and 19 for non-line-of-sight (NLOS) (each of them has 20 MPCs) scenarios, our results in both frequency bands appear to have smaller number of clusters and MPCs per cluster.

As far as the relation between the cluster power and cluster propagation distance is concerned, Fig. 10 shows the empirical data obtained from the measurements and our clustering process. Linear regression fits well with the empirical data in both bands. The fitting model can be expressed as

$$P_c = A \log_{10}(d_c) + B, \quad (2)$$

where $P_c$ and $d_c$ are the cluster power in dB and cluster propagation distance in m, respectively. A simple liner regression provides that $(A, B)$ is equal to $(-30.5, -58.0)$ in the 28-GHz band and equal to $(-24.8, -78.1)$ in the 140-GHz band.

TABLE V
MEAN AND STANDARD DEVIATION OF THE NUMBER OF CLUSTERS $N$ AND THE NUMBER OF MPCs PER CLUSTER $M$ FOR SHOPPING MALL SCENARIO.

| Parameter | | 28-GHz band | 140-GHz band |
|---|---|---|---|
| $N$ | $\mu$ | 7.9 | 5.9 |
| | $\sigma$ | 3.6 | 2.1 |
| $M$ | $\mu$ | 5.4 | 3.8 |
| | $\sigma$ | 6.0 | 2.5 |

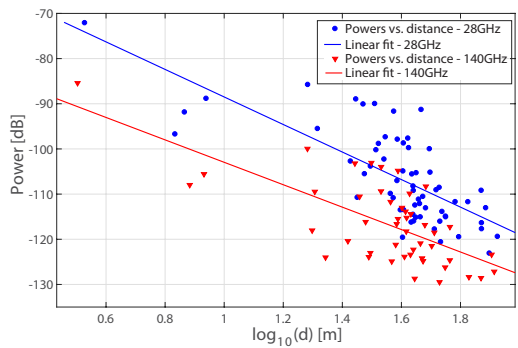

Fig. 10. Empirical cluster power [dB] versus cluster distance [$log_{10}$ (m)], and the linear fit at 28 and 140 GHz.

TABLE VI
MODEL PARAMETERS AND FITTING ERROR FOR THE COMPOSITE PAS IN AZIMUTH OF ALL CLUSTERS.

| Model | | 28-GHz band | 140-GHz band |
|---|---|---|---|
| Gaussian | $\sigma$ [°] | 17.9 | 18.0 |
| | RMSE | 0.011 | 0.005 |
| Von Mises | $\kappa$ [°] | 8.2 | 8.2 |
| | RMSE | 0.086 | 0.085 |

The generation of the cluster offset angle in 3GPP is based on the distribution of the composite power angular spectrum (PAS), and the AoAs can be determined via inverse Gaussian [4] or inverse wrapped Gaussian functions. The latter can be closely approximated by Von Mises distribution that is mathematically simpler and more tractable. The fitting models for the normalized cluster power $P_n/\max(P_n)$ versus offset AoA $\phi_n$ to both Gaussian and Von Mises distributions are

$$\frac{P_n}{\max(P_n)} = \exp[-(\phi_n/\sigma)^2], \quad (3)$$

$$\frac{P_n}{\max(P_n)} = \frac{e^{\kappa \cos \phi_n}}{2\pi I_0(\kappa)}, \quad (4)$$

respectively, where $I_0(\kappa)$ is the modified Bessel function of order 0, $1 \leq n \leq N$ is the clustering index, $N$ is the total number of clusters. The results in Table VI show that the model parameters are similar between 28 and 140 GHz bands.

ACKNOWLEDGEMENT

The research leading to the results presented in this paper received funding from Nokia Bell Labs, Finland.